# UNDERWATER IMAGE ENHANCEMENT USING CONVOLUTIONALNEURAL NETWORK

**Anushka Yadav, Mayank Upadhyay, Ghanapriya Singh**

[1] *National Institute of Technology, Department of Electronics Engineering, Uttarakhand*
*ghanapriya@nituk.ac.in*

*Abstract*

This report proposes a method for underwater image enhancement using the principle of histogram equalization. Since underwater images have a global strong dominant colour, their colourfulness and contrast are often degraded. Before applying the histogram equalisation technique on the image, the image is converted from coloured image to a gray scale image for further operations. Histogram equalization is a technique for adjusting image intensities to enhance contrast. The colours of the image are retained using a convolutional neural network model which is trained by the datasets of underwater images to give better results.

.

*Keywords:* Underwater image enhancement, 1-D CNN, 2-D CNN, LSTM

___________

*\*Corresponding author:* ghanapriya@nituk.ac.in

# INTRODUCTION

Image enhancement is the process of adjusting digital images so that the results are more suitable for display or further image analysis. Underwater photography, underwater imaging has also been an important source of interest in different branches of technology and scientific research, such as inspection of underwater infrastructures and cables, detection of manmade objects, control of underwater vehicles, marine biology research, and archaeology. Different from common images, underwater images suffer from poor visibility resulting from the attenuation of the propagated light, mainly due to absorption and scattering effects. The absorption substantially reduces the light energy, while the scattering causes changes in the light propagation direction. They result in foggy appearance and contrast degradation, underwater image normally has several problems including limited range of visibility, low contrast, non-uniform lighting, bright artifacts, noise, blurring, and diminishing colour. It is a challenging task to remove a dominant color while keeping sharpness or brightness for various image processing tasks, e.g., image recognition. Consequences from these problems, researchers are improving the image contrast to extract as many information as possible by applying various algorithms. Different from common images, underwater images suffer from poor visibility resulting from the attenuation of the propagated light, mainly due to absorption and scattering effects. The absorption substantially reduces the light energy, while the scattering causes changes in the light propagation direction. They result in foggy appearance and contrast degradation.

The proposed image enhancement is based on three main steps, first step is decolouring the image. The first step includes conversion of a RGB image to gray scale image so that instead of dealing with three planes and their values, major processing will be done on the intensity values in the second step. The second step of this process is performing histogram equalisation on the converted gray scale image, by which the layer of a particular intensity that was shadowing over the entire image hiding the necessary details is removed. This histogram equalised greyscale image is converted back into a coloured RGB image using a convolutional neural network model. The model is trained with Dataset containing underwater images.

# LITERATURE REVIEW

Based on the research done by Tatsuya Baba, Keishu Nakamura, Seisuke Kyochi, and Masahiro Okuda cited in the research paper [1], he described a novel image enhancement method for underwater images based on discrete cosine eigenbasis transformation (DCET). Since underwater images have a global strong dominant colour, their colourfulness and contrast are often degraded. Typical colour correction methods for natural images, i.e., computational colour constancy, achromatize the coloured illuminant of input images by dividing with an estimated illuminant colour. However, this procedure produces unwanted colour artifacts for underwater images. To solve this problem, we introduce a novel assumption that achromatic illuminant images should have the DCT basis vectors as their principal component vectors, which is termed as discrete cosine eigenbasis (DCE). According to the assumption, we achromatize underwater images by using a DCET that transforms the input image to the images having the DCE. By incorporating post image enhancement techniques, the proposed method provides sharper and brighter visual quality than conventional colour correction and image enhancement methods. This paper proposed the colour correction and image enhancement method for underwater images. Based on the new assumption, we introduced the DCET in the CCC procedure, which enables us to robustly correct the dominant colour of underwater images. In the image enhancement, a chroma contrast enhancement image and a luma contrast enhancement image are generated, and an image having a vivid colour and sharp detail is generated through the integration. The experiments showed the effectiveness of the proposed method by comparing with the conventional methods.

Based on the research done by Ahmad Shahrizan Abdul Ghani, Nor Ashidi Mat Isa cited in the research paper[1], they described the major issues with underwater images. The physical property of water medium causes attenuation of light travels through the water medium, resulting in low contrast, blur,inhomogenous lighting and colour diminishing of the underwater images. This paper extends the method of enhancing the quality of underwater image. The proposed method in this research consists of two stages. at the first stage, the contrast correction technique is applied to the image, where the image is applied with the modified Von Kries hypothesis and stretching the image into two different intensity image at the average value with respect to the Rayleigh distribution. At the second stage , colour correction technique is applied to the image where the image is first converted into hue- saturation value(HSV) colour model. The modification of the colour component increases the image colour performance. To evaluate the proposed technique 300 underwater images, which are captured in three different Malaysian islands namely Tioman, Langkawi and Perhention are used in the experiments. The proposed technique is compared with ICM and UCM in terms of entropy, mean square error (MSE), and peak signal to noise ratio (PSNR). Two other methods are also used for comparison, namely, the pixel distribution shifting colour correction (PDSCC). PDSCC is the latest contrast enhancement technique that is also designed for underwater images.

The results of our method has been compared to the above mentioned methods and it has been observed that on comparing with these proposed method our method gives better results based on the entropy, mean square error(MSE) and peak to signal noise ratio.

## METHODOLOGY

The proposed image enhancement is based on three main steps, first step is decolouring the image. Fig 1 shows the process of proposed method. In the first step, the coloured image which is specifically dominant in a few colours that are likely to be blue and green is converted into a grayscale image. The first step includes conversion of a RGB image to grayscale image so that instead of dealing with three planes and their values, major processing will be done on the intensity values in the second step. The second step of this process is performing histogram equalisation on the converted gray scale image, by which the layer of a particular intensity that was shadowing over the entire image hiding the necessary details is removed. In case of the coloured image it appears to be the dominant colour region (a layer of blue or green colour). After performing the histogram equalisation, the image appears to hold required details which were not present at first. This histogram equalised grayscale image is converted back into acoloured RGB image using a convolutional neural network model. The model is trained with Dataset containing underwater images.

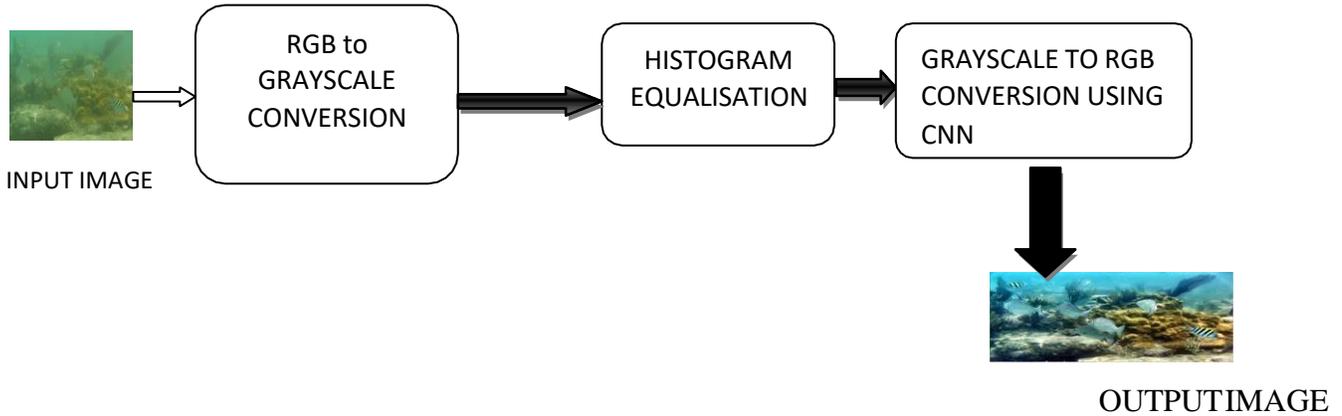

Fig1: Flowchart of the Methodology

**Histogram equalization**:

It is a method in image processing of contrast adjustment using the image's histogram. This method usually increases the global contrast of many images, especially when the usable data of the image is represented by close contrast values. Through this adjustment, the intensities can be better distributed on the histogram. This allows for areas of lower local contrast to gain a higher contrast. Histogram equalization accomplishes this by effectively spreading out the most frequent intensity values.

Let f be a given image represented as a mr by mc matrix of integer pixel intensities ranging from 0 to L − 1. L is the number of possible intensity values, often 256. Let p denote the normalized histogram of f with a bin for each possible intensity. So

$$p_n = \frac{\text{number of pixels with intensity } n}{\text{total number of pixels}} \quad n = 0, 1, ..., L-1.$$

The histogram equalized image g will be defined by

$$g_{i,j} = \text{floor}\left((L-1) \sum_{n=0}^{f_{i,j}} p_n\right), \quad (1)$$

where floor() rounds down to the nearest integer. This is equivalent to transforming the pixel intensities, k, of f by the function

$$T(k) = \text{floor}\left((L-1) \sum_{n=0}^{k} p_n\right).$$

The motivation for this transformation comes from thinking of the intensities of f and g as continuous random variables X, Y on [0, L − 1] with Y defined by

$$Y = T(X) = (L-1) \int_0^X p_X(x)dx, \qquad (2)$$

where pX is the probability density function of f. T is the cumulative distributive function of X multiplied by (L − 1). Assume for simplicity that T is differentiable and invertible. It can then be shown that Y defined by T(X) is uniformly distributed on [0, L − 1], namely that

$$p_Y(y) = \tfrac{1}{L-1}.$$

$$\int_0^y p_Y(z)dz = \text{probability that } 0 \leq Y \leq y$$
$$= \text{probability that } 0 \leq X \leq T^{-1}(y)$$
$$= \int_0^{T^{-1}(y)} p_X(w)dw$$

$$\frac{d}{dy}\left(\int_0^y p_Y(z)dz\right) = p_Y(y) = p_X(T^{-1}(y))\frac{d}{dy}(T^{-1}(y)).$$

Note that $\frac{d}{dy}T(T^{-1}(y)) = \frac{d}{dy}y = 1$, so

$$\frac{dT}{dx}\Big|_{x=T^{-1}(y)}\frac{d}{dy}(T^{-1}(y)) = (L-1)p_X(T^{-1}(y))\frac{d}{dy}(T^{-1}(y)) = 1,$$

which means $p_Y(y) = \tfrac{1}{L-1}$.

Our discrete histogram is an approximation of pX (x) and the transformation in Equation 1 approximates the one in Equation 2. While the discrete version won't result in exactly flat histograms, it will flatten them and in doing so enhance the contrast in the image.

The grayscale image is then converted to coloured image by using convolutional neural network.

## Convolutional Neural Network:

In deep learning, a convolutional neural network (CNN) is a class of deep neural networks, most commonly applied to analyzing visual imagery. CNNs use a variation of multilayer perceptrons designed to require minimal pre-processing. They are also known as shift invariant or space invariant artificial neural networks (SIANN), based on their shared-weights architecture and invariance characteristics. Convolutional networks were inspired by biological processes in that the connectivity pattern between neurons resembles the organization of the animal visual cortex. Individual cortical neurons respond to stimuli only in a restricted region of the visual field known as the receptive field. The receptive fields of different neurons partially overlap such that they cover the entire visual field. CNNs use relatively little pre-processing compared to other image classification algorithms. This means that the network learns the filters that in traditional algorithms were hand-engineered. This independence from prior knowledge and human effort in feature design is a major advantage.

## Autoencoder :

An autoencoder learns to compress data from the input layer into a short code, and then uncompress that code into something that closely matches the original data. This forces the autoencoder to engage in dimensionality reduction, for example by learning how to ignore noise. Some architectures use stacked sparse autoencoder layers for image recognition. The first encoding layer might learn to encode easy features like corners, the second to analyze the first layer's output and then encode less local features like the tip of a nose, the third might encode a whole nose, etc., until the final encoding layer encodes the whole image into a code that matches (for example) the concept of "cat". The decoding layers will learn to decode the learnt code representation back into its original form as close as possible.

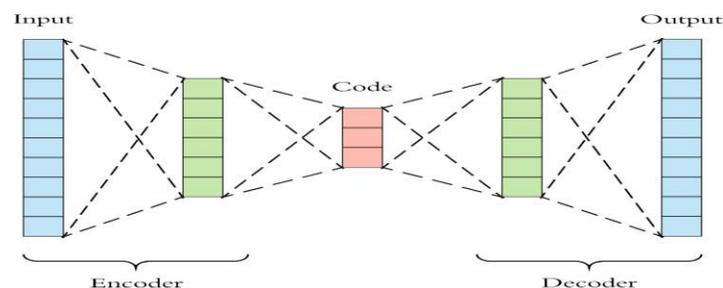

# RESULT AND CONCLUSION

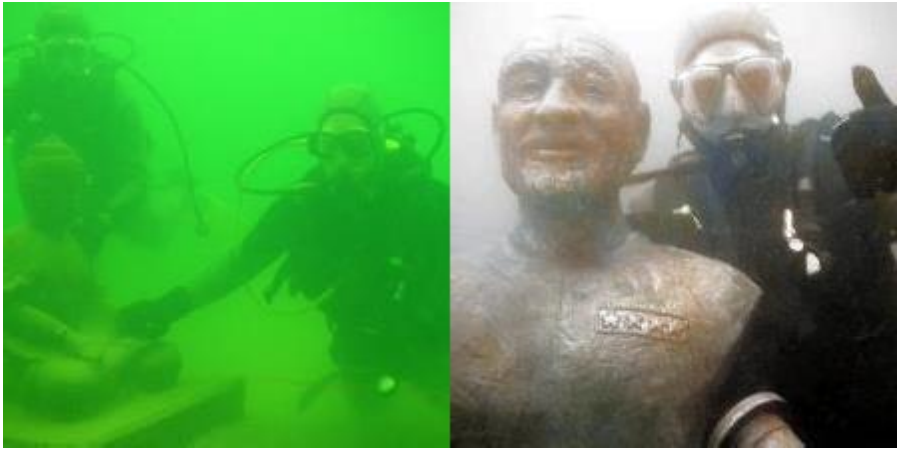

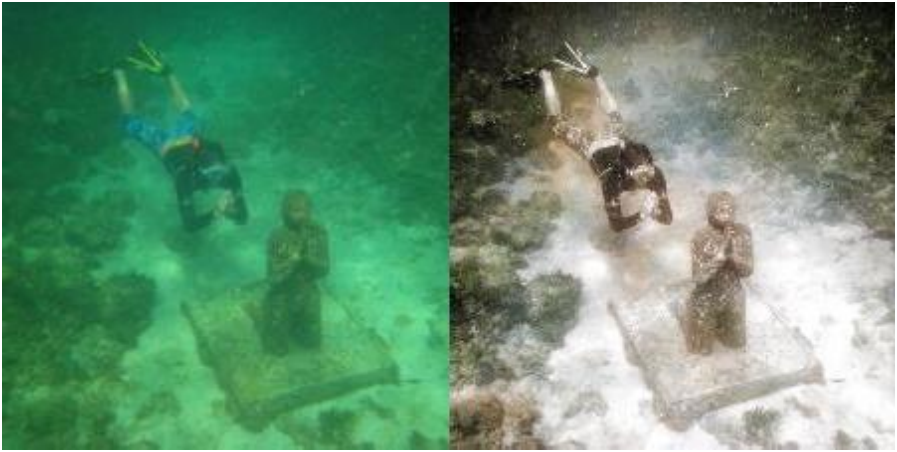

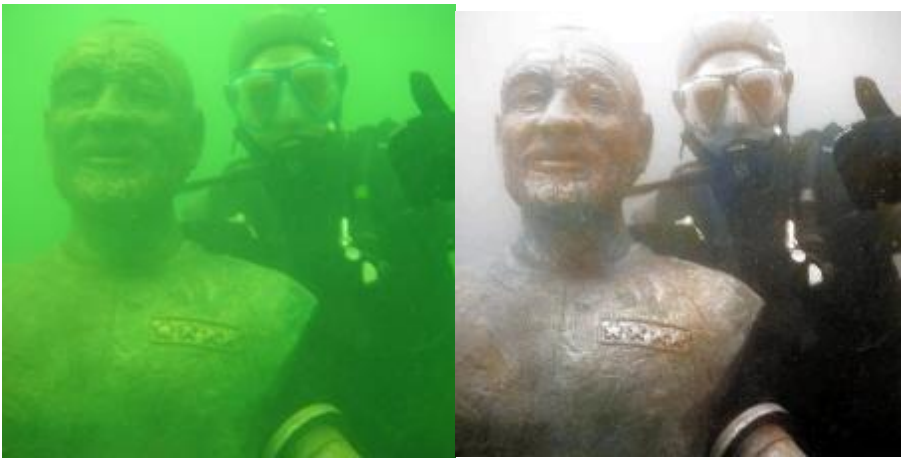

**Input**                                          **Output**

# **CONCLUSION :**

We have presented an alternative approach to enhance underwater images. Our strategy builds on the histogram equalisation and does not require additional information than the single original image. We have shown in our experiments that our approach is able to enhance a wide range of underwater images (e.g. different cameras, depths, light conditions) with high accuracy, being able to recover important faded features and edges. Moreover, for the first time, we demonstrate the utility and relevance of the proposed image enhancement technique for several challenging underwater computer vision napplications. The proposed methods produces lesser Mean Square Error (MSE), better Entropy and Peak Signal to Noise Ratio (PSNR) value when compared to existing ICM techniques.